\documentclass[prl,twocolumn,aps,amssymb,showpacs]{revtex4} 

\usepackage{times}

\begin{document}

\title{Transients due to instabilities hinder Kardar-Parisi-Zhang scaling: 
a unified derivation for surface growth by electrochemical and chemical 
vapor deposition}

\author{Rodolfo Cuerno}
\affiliation{Departamento de Matem\'aticas \& GISC, Universidad
Carlos III de Madrid, Avenida Universidad 30, 28911 Legan\'{e}s, Spain}

\author{Mario Castro}
\affiliation{Universidad Pontificia de Comillas, 28015 Madrid, Spain,}

\affiliation{Departamento de F\'{\i}sica de Materiales \& GISC, 
Facultad de Ciencias F\'{\i}sicas, U.\ Complutense de Madrid, 28040 Madrid, 
Spain}

\date{{\em To appear in Physical Review Letters, December 3, 2001}}

\begin{abstract}
\vspace*{0.5cm}
We propose a unified moving boundary problem for surface growth by
electrochemical and chemical vapor deposition, which is
derived from constitutive equations into which stochastic
forces are incorporated.
We compute the coefficients in the interface equation of
motion as functions of phenomenological parameters. The equation features
the Kardar-Parisi-Zhang (KPZ) non-linearity and instabilities which,
depending on surface kinetics, can hinder the asymptotic KPZ scaling.
Our results account for the universality and the experimental scarcity of
KPZ scaling in the growth processes considered.
\end{abstract}

\pacs{68.35.Ct, 64.60.Ht, 81.15.Gh, 81.15.Pq}

\maketitle

The dynamics of rough surfaces \cite{alb+krug} is a subject of high interest. 
This is due both to its implications for processes of technological 
relevance \cite{ecd,cvd}, and to the interesting instances that 
it offers of extended systems evolving in the presence of fluctuations 
\cite{os}. A very successful framework for the study of rough interfaces has 
been the use of stochastic growth equations for the interface height. 
Among these, the one proposed by Kardar, Parisi and Zhang (KPZ) \cite{kpz} 
has played a prominent role, since in particular it has enabled 
connections to be made with other physical problems, 
like directed polymers in disordered media or randomly stirred fluids
\cite{alb+krug}. On the basis of a coarse-grained description of 
surface growth and symmetry arguments, the KPZ equation was 
initially expected to describe the dynamics of surfaces growing, $e.g.$ at the 
expense of a vapor phase,
in the absence of conservation laws, and is thus expected to be relevant to 
such diverse physical growth systems as electrochemical deposition (ECD) 
\cite{ecd} or chemical
vapor deposition (CVD) \cite{cvd}. The generality of KPZ scaling
would be a consequence of the phenomenon of {\em universality} observed for 
the scaling properties of rough surfaces. However, to date very few experiments
have been reported which are compatible with the predictions of the KPZ 
equation \cite{kpzexp,schilardi,ourcvd}. Moreover, mere symmetry arguments 
do {\em not} enable a detailed connection with 
phenomenological parameters describing specific experimental systems, while 
detailed {\em derivations} of the KPZ equation were achieved for discrete or
continuous theoretical models \cite{deriv}, only indirectly related with 
experiments. These facts have led to invoking additional effects on the same 
coarse-grained level, such as specific noise statistics, non-local effects,
etc.\ \cite{alb+krug}, in order to 
account for the difference between the observed and the predicted scaling 
behaviors of rough surfaces. However, a wide range of scaling exponents ensued,
there being no theoretical argument that could identify the correct
exponents for a specific growth experiment. 

In this Letter we study two of the main techniques employed in experiments 
on non-conserved surface growth, namely ECD and CVD. 
These techniques have actually
played a preeminent role in the study of pattern formation 
\cite{revpatt}, but only recently have they been
shown to provide experimental realizations of rough interfaces in the KPZ 
universality class \cite{schilardi,ourcvd}. The asymptotic behavior is in both
systems preceded by exceedingly long unstable transients, extending in $e.g.$ 
the experiments in \cite{ourcvd} for up to two days deposition time. 
This complex time behavior is thus far unaccounted for 
on general grounds by any coarse-grained 
continuum model. Here we start from the constitutive equations of ECD and
CVD, into which we allow for stochastic forces, following a similar
treatment to that employed in studies of solidification 
\cite{karma}, step dynamics \cite{km+plm} or fluid imbibition \cite{dube+barna}.
We show that ECD and CVD can both be described 
within a unified framework, which provides a stronger statement on 
universality in non-conserved growth phenomena than that restricted to
scaling behavior.
Moreover, we compute the coefficients appearing in the 
ensuing stochastic interface equation of motion (IEOM) as functions of
the phenomenological parameters characterizing the 
corresponding physical growth process.
The IEOM features the expected KPZ 
nonlinearity, but also instabilities which can hinder asymptotic KPZ scaling. 
Specifically, for the case of non-instantaneous growth events at the surface, 
the IEOM is a stochastic 
generalization \cite{km+plm,erosion} of the Kuramoto-Sivashinsky equation 
\cite{ksorig}, for which very long transients due to instabilities are known 
to occur \cite{kstransients} before scaling behavior can be observed. 
Thus, there is no need to invoke additional effects at a coarse-grained 
level in order to account for the difficulty in observing KPZ scaling, but 
rather they are due to the long unstable transients which will quite 
generically occur. 
Our approach also accounts for features of discrete 
growth models \cite{mbdla} and our conclusions are thus expected to apply 
rather generally for non-conserved growth systems.

We first consider growth by CVD. A successful model of this type of growth 
was formulated and developed in \cite{brekel+pg+brz}. 
A stagnant diffusion layer of infinite vertical extent 
is assumed to exist above the substrate upon 
which an aggregate will grow. Particles of an intermediate species 
$[$concentration $c(x,z,t) \equiv c({\bf r},t)$, where $x$ is the coordinate
along the initial one-dimensional substrate and $z$ is the growth
direction$]$ diffuse through the 
stagnant layer. When they meet the surface they react in order to stick
to the aggregate, this occurring with an efficiency measured by a kinetic 
mass-transfer coefficient $k_D$ \cite{stick}. 
Additional curvature driven effects, such as 
surface diffusion and evaporation-condensation,
can influence the local growth velocity of the aggregate. Moreover, we will 
take into account local fluctuations in the vapor phase and surface 
diffusion currents, as well as in the deposition events, 
in order to account for the experimental relevance of fluctuations for the 
morphology of surfaces grown by CVD \cite{ourcvd}.
We thus propose the following stochastic generalization of the deterministic 
model of CVD
\cite{brekel+pg+brz}:
\begin{eqnarray}
& & \frac{\partial c}{\partial t}=D\nabla^2 c -\nabla\cdot{\bf q}, 
\label{ecdif}\\
& & k_D(c-c_{eq}^0-\Gamma\kappa + \chi)= (D\nabla c - {\bf q}) 
\cdot {\bf n} , \label{ecstick} \\
& & {\bf v}\cdot{\bf n}=\Omega (D\nabla c - {\bf q}) \cdot{\bf n} - 
B \nabla^2_{\rm s}\kappa - \Omega \nabla \cdot {\bf p}, \label{ecvel} \\
& & c(x,z \rightarrow\infty,t)=c_a . \label{ecinfty}
\end{eqnarray}
In Eq.\ (\ref{ecdif}), $D$ is the diffusion constant, and the conserved noise 
${\bf q}({\bf r},t)$ represents fluctuations in the concentration associated 
with diffusion through the stagnant layer. 
Eqs.\ (\ref{ecstick}) and (\ref{ecvel}) 
hold at any point on the aggregate surface ${\bf r} = {\bf s}$, the symbol 
$\nabla^2_{\rm s}$ denoting the surface Laplacian operator and
${\bf n}$ the local unit normal vector. 
Eq.\ (\ref{ecstick}) is a mixed boundary condition which relates the 
diffusion current arriving at the aggregate from the stagnant layer with 
the material wich actually deposits, via the kinetic coefficient $k_D$. 
The noise term $\chi$ represents fluctuations in the deposition events 
\cite{km+plm}, $c^0_{eq}$ is the equilibrium concentration for a flat inteface,
$\kappa$ is the local mean curvature, and $\Gamma = \gamma c^0_{eq} 
\Omega/(k_B T)$, with  $\gamma$ the surface tension ---whose anisotropy will 
be neglected, $i.e.$ we will consider an amorphous or polycrystalline 
aggregate---, and $T$ temperature.  
In Eq.\ (\ref{ecvel}), ${\bf v}$ is the local aggregate velocity,
$\Omega$ is the atomic volume of the depositing species, and the coefficient
of the surface diffusion current \cite{alb+krug} $B = D_{\rm s} \nu_{\rm s} 
\gamma \Omega^2/ k_B T$, with $D_{\rm s}$ the surface diffusivity and  
$\nu_{\rm s}$ the surface concentration of particles. 
Moreover, ${\bf p}$ is a noise term associated with the surface diffusion
current. Finally, $c_a$ in (\ref{ecinfty}) is a constant value held fixed at 
the edge of the stagnant layer. 
Note that the deterministic model of CVD \cite{brekel+pg+brz} is 
recovered by neglecting ${\bf q}$, ${\bf p}$, and $\chi$ in 
(\ref{ecdif})-(\ref{ecinfty}). 
We consider zero-mean, uncorrelated and white noise terms. 
A local equilibrium hypothesis \cite{karma,km+plm} then allows 
us to determine their variances to be \cite{us}
\begin{eqnarray}
\langle q^i({\bf r},t)q^j({\bf r}^\prime,t^\prime)\rangle&=&
2D c({\bf r},t) \delta_{ij}
\delta({\bf r}-{\bf r}^\prime)\delta(t-t^\prime),\label{Q_def} \\
\langle p^i({\bf s},t)p^j({\bf s}^\prime,t^\prime)\rangle&=& 
2D_{\rm s} \nu_{\rm s} \delta_{ij}
\delta({\bf s}-{\bf s}^\prime)\delta(t-t^\prime),\label{P_def} \\
\langle \chi({\bf s},t)\chi({\bf s}^\prime,t^\prime)\rangle&=&
(2 c({\bf s},t)/k_D) \delta({\bf s}-{\bf s}^\prime)\delta(t-t^\prime).
\label{I_def}
\end{eqnarray} 

Before studying the interface dynamics predicted by model 
(\ref{ecdif})-(\ref{ecinfty}), let us show that it also describes 
electrochemical deposition, under a proper interpretation of the fields 
and parameters appearing.
For simplicity, we assume a rectangular thin cell in which the two electrodes 
are made of the same metal, the cell being filled with a {\em dilute} 
solution of a salt of this metal. In a growth experiment by ECD \cite{ecd}, 
an electric field is applied driving the motion of cations towards the 
cathode, whereupon they stick via a reduction reaction,
leading to the growth of an aggregate.
The constitutive equations, neglecting convection of the electrolyte,
\cite{chazalviel} describe diffusion of cations (concentration $C$) and of 
anions (concentration $A$), together with Poisson's equation for the electric
field accross the cell. This highly non-trivial system can be somewhat 
simplified under the electroneutrality condition \cite{ecd2} 
$z_a A = z_c C$, where $ez_c$ and $-ez_a$ are the cationic and anionic charges,
implying
\begin{equation}
\partial_tC=D\nabla^2C , \label{difecd}
\end{equation}
where $D=(\mu_cD_a+\mu_aD_c)/(\mu_a+\mu_c)$ is the ambipolar diffusion 
constant, $\mu_{c,a}$ being the cationic and anionic mobilities.
The electric field configuration and the anion dynamics are implicit in the
definition of $D$ and in the boundary conditions (BC), which we need to specify.
The simplest BC are that the anion flux be zero both at the anode and at the 
cathode, where only cations contribute to the aggregate growth. The current 
density at the cathode surface is then \cite{ecd}
\begin{equation}
{\bf J} \cdot {\bf n} = - [z_c D_c F/(1-t_c)] \nabla C \cdot {\bf n},
\label{corr1}
\end{equation}
where $t_c \equiv \mu_c/(\mu_a+\mu_c)$ and $F$ is Faraday's constant.
Moreover, charge transport at the cathode is an activated process, whose 
balance is described by the Butler-Volmer equation \cite{ecd,km}
\begin{equation}
J=J_0\left[ e^{\frac{(1-\beta)\eta z_cF}{RT}}-
e^{\frac{-(\beta \eta + \eta_s) z_cF}{RT}} C/C_a \right],
\label{BV}
\end{equation}
where $J_0$ is the exchange current density in equilibrium, $\beta$ is a 
coefficient between 0 and 1 describing the asymmetry of the 
energy barrier related to the cation reduction reaction, $C_a$ is the 
initial cation concentration, $R$ is the 
gas constant, and $\eta$ is the overpotential, from which a surface 
curvature contribution $\eta_s$ has been singled out. 
By defining the concentration field $c \equiv D_c C/[D (1-t_c)]$ 
($c_a$ and $c^0_{eq}$ are defined accordingly) and performing an expansion 
of Eq.\ (\ref{BV}) for a small value of 
$\eta_s = \Omega^{\scriptscriptstyle ECD}
\gamma^{\scriptscriptstyle ECD} R \kappa / (z_c F k_B)$ \cite{ecd}, Eqs.\ 
(\ref{ecdif})-(\ref{ecinfty}) provide the stochastic generalization of the 
ECD model (\ref{difecd})-(\ref{BV}) incorporating surface tension and surface 
diffusion effects at the aggregate surface.
Note that Eqs.\ (\ref{corr1}) and (\ref{BV}) together 
amount to a mixed BC \cite{hill} on $c$ of the type of Eq.\
(\ref{ecstick}), with a kinetic coefficient $k_D^{\scriptscriptstyle ECD} 
\equiv J_0 e^{-\beta z_c F \eta/RT} D (1-t_c)/(z_c F C_a D_c)$. 
For instance, the $\eta \rightarrow - \infty$ limit of completely efficient 
reduction at the cathode leads to an absorbing boundary condition there of 
the type $c= c^0_{eq} + \Gamma \kappa$. 

We are now in a position to study the generic model 
(\ref{ecdif})-(\ref{ecinfty}) and draw conclusions for 
the two diverse growth systems considered.
We follow a similar approach to that in
\cite{karma,km+plm,dube+barna} and references 
therein. Namely, we first note that in the zero noise limit
Eqs.\ (\ref{ecdif})-(\ref{ecinfty}) support 
a flat solution $c({\bf r},t) = c(z,t)$,
moving at a constant velocity $V= k_D [\Omega (c_a - c^0_{eq})-1]$.
By the use of the diffusion Green function, we project the moving boundary 
problem onto the aggregate surface. Finally, we perform a perturbation 
expansion and a long wavelength analysis in order to derive a stochastic 
differential equation for a local deviation $\zeta(x,t)$ (in the frame moving 
with velocity $V$) from the flat interface solution. Details will be 
reported elsewhere \cite{us}. The results are conveniently
classified by the value of the kinetic coefficient $k_D$ \cite{frankel}.

\medskip

\noindent
{\em Instantaneous surface kinetics ($k_D \rightarrow \infty$)}.--- \\
Denoting by $\zeta_k(t)$ the $k$-th Fourier mode of $\zeta(x,t)$, 
in the case of infinitely fast reaction kinetics at the interface
(absorbing boundary condition) the IEOM reads 
\begin{equation}
\partial_t \zeta_k(t) = \omega(k) \, \zeta_k(t) + \frac{V}{2}
{\cal F}_k[(\nabla \zeta)^2] + \beta_k^{\rm absorb}(t) .
\label{1aec}
\end{equation}
Here ${\cal F}_k[f(x)]$ denotes the $k$-th Fourier mode 
of $f(x)$, the non-linear term in (\ref{1aec}) having the expected KPZ form,
and $\beta_k^{\rm absorb}(t)$ is an 
{\em additive} noise term whose correlations depend on the dispersion relation 
$\omega(k)$ \cite{corrs}. Note that, in principle, the system 
(\ref{ecdif})-(\ref{ecinfty}) has {\em multiplicative} noise. However,
similarly to \cite{km+plm}, to lowest non-linear order in $\zeta(x,t)$
the IEOM features only {\em additive} noise terms. For Eq.\ (\ref{1aec})
the dispersion relation reads 
$\omega(k) = V |k| (1-d_0 l_D k^2) [1- d_0/l_D + (d_0^2/4 - B/D) k^2]^{1/2}
+ D (d_0^2/2 - B/D) k^4 - 3d_0 D k^2/l_D$, where we have 
defined a capillarity length $d_0 = \Gamma \Omega$ and a diffusion length
$l_D = D/V$. As we see, due to the shape of $\omega(k)$,
the IEOM (\ref{1aec}) is non-local in space. This is a 
reflection of the diffusional instabilities present in the system
\cite{revpatt}. 
For instance, in the absence of surface diffusion currents ($B =0$), and 
as long as $d_0 \ll l_D$, the dispersion relation of Eq.\ (\ref{1aec}) is of 
the Mullins-Sekerka type \cite{revpatt}, 
$\omega(k) \simeq V |k| (1- d_0 l_D k^2)$.
However, the IEOM does have additive noise and a KPZ non-linear term with 
coefficient $V/2$, both facts as expected on general grounds \cite{alb+krug}.
Although the behavior of Eq.\ (\ref{1aec}) at large scales is not completely
known, both a scaling argument and preliminary numerical simulations
indicate that KPZ scaling is not asymptotic under these growth conditions 
\cite{tesis}. 

\medskip

\noindent
{\em Non-instantaneous surface kinetics ($k_D < \infty$)}.--- \\
For the case of non-instantaneous deposition events at the surface, the 
dispersion relation $\omega(k)$ turns out to depend only on even powers 
of $k$, which allows to express the IEOM directly in configuration space,
featuring only {\em local} terms [we omit the ${\bf r}$ and $t$ dependencies of 
$\zeta({\bf r}, t)$]:
\begin{equation}
\partial_t \zeta = -a_2 \nabla^2 \zeta -
a_4 \nabla^4 \zeta + \frac{V}{2} (\nabla \zeta)^2 + 
\beta^{\rm mixed}({\bf r},t) , \label{2aec}
\end{equation}
where $a_2 = k_D l_D \Delta$, $a_4 = k_D l_D^2 d_0 \Delta /[1-(d_0/l_D)^{1/2}]
\\ + B (1+k_D/V)$, and $\Delta = 1 - d_0/l_D$. Again,
$\beta^{\rm mixed}({\bf r},t)$ is an {\em additive} noise with 
$\omega(k)$-dependent correlations \cite{corrs}.
As in Eq.\ (\ref{1aec}), the coefficient of the KPZ-nonlinearity is $V/2$.
However, while (\ref{1aec}) always has a band of linearly unstable modes,
this only happens in Eq.\ (\ref{2aec}) if $\Delta > 0$, $i.e.$, when 
surface tension is unable to counteract the diffusional instabilities
($d_0 < l_D$). In this unstable case, (\ref{2aec}) is the stochastic
Kuramoto-Sivashinsky (KS) equation, already encountered in other interface 
dynamics contexts, such as step dynamics \cite{km+plm}, or surface erosion by 
ion-beam sputtering \cite{erosion}. In the stochastic KS system, there
is a linearly most unstable mode $k_m = (2 l_D d_0)^{-1/2}$ 
(assuming $B =0$), whose onset time is $\omega_m^{-1}
= 2/(k_D l_D k_m^2)$. For asymptotic times, KPZ scaling is obtained, but only
after an exceedingly long transient \cite{kstransients}. 

From the above analysis of Eqs.\ (\ref{1aec}) and (\ref{2aec}), 
KPZ scaling should be expected only for slow surface 
kinetics, and will nevertheless be affected by early time instabilities, 
unless the capillary length is larger than the typical diffusion length in the 
system. KPZ scaling corresponds to the {\em conformal growth} mode, which is 
most interesting for applications of CVD-grown films and can indeed be 
achieved under industrial conditions by tuning the relative values 
of $k_D$, $l_D$, and $d_0$ \cite{cvd,ourcvd,brekel+pg+brz}. 
In the case of ECD, we can further verify the predictions from model 
(\ref{ecdif})-(\ref{ecinfty}) via Eqs.\ (\ref{1aec}), (\ref{2aec}) with 
both experimental \cite{schilardi,pr} and Monte-Carlo \cite{mbdla} studies.
For instance, in the ECD experiment in \cite{pr} the 
diffusion length $l_D \approx 2$ cm is the largest length scale in the system,
and a Mullins-Sekerka dispersion relation is reported; 
consequently, after a stable transient 
the diffusional instabilities that occur completely override any scaling 
behavior at long times. As a difference, in the experiments of 
\cite{schilardi} the growing aggregate undergoes an unstable transient,
beyond which its surface stabilizes into the KPZ stationary 
state. This behavior is qualitatively compatible with that of the noisy 
KS equation: from the experimental value of the branch spacing one has
$k_m \simeq 1.3 \times 10^3$ cm$^{-1}$. Using \cite{schilardi}
$D \simeq 10^{-5}$ cm$^2$ s$^{-1}$ and 
$V \simeq 2 \times 10^{-4}$ cm s$^{-1}$,
one obtains a typical value $d_0 = 1/(2 l_D k_m^2) \simeq 5 \times 10^{-6}$ cm,
hence indeed $d_0 \ll l_D \simeq 0.05$ cm. 
Moreover, the instability occurs after $1/\omega_m \simeq 0.3 \times 10^3$ s, 
which allows to estimate $k_D \simeq 8 \times 10^{-8}$ cm s$^{-1} \ll V$,
hence conditions are in the slow kinetics regime.

Eqs.\ (\ref{ecdif})-(\ref{ecinfty}) also apply to the 
dynamics of discrete growth models such as the MBDLA model \cite{mbdla}, which
is a generalization of the DLA model \cite{alb+krug,revpatt} to the case of a 
finite concentration of random walkers performing biased random walks (bias 
parameter $p$), which stick to the growing aggregate with a finite sticking 
probability, $s$. MBDLA reproduces ECD experiments for one-dimensional 
substrates {\em quantitatively} \cite{mbdla}, and ECD experiments for 
two-dimensional substrates {\em qualitatively} \cite{copper}. 
Thus, the sticking probability $s$ plays the role of 
a noise reduction parameter, in the sense that for small $s$ values
the system reaches faster its KPZ asymptotic scaling behavior; this role 
is played in our continuum model by the kinetic mass transfer coefficient
$k_D$ \cite{stick}. Furthemore, the bias $p$ is proportional to the aggregate 
velocity $V$. As the characteristic branch spacing is, according to 
(\ref{2aec}), approximately equal to $(l_D d_0)^{-1/2}$, the continuum model 
predicts a ramified-to-compact transition as $p$ increases, such as is 
observed in MBDLA \cite{mbdla}. Moreover, since the coefficient of the 
KPZ term also increases, the corresponding scaling is expected to occur 
earlier as $V$ (or $p$) increases, again as is observed in the discrete 
model.   

In summary, our study of CVD and ECD leads us to expect diffusional 
instabilities to generically hinder KPZ scaling in non-conserved growth 
experiments. Our continuum approach from constitutive equations allows to 
perform detailed comparison with phenomenological parameters in experiments, 
and also provides a physical interpretation for features of discrete models of 
kinetic roughening such as noise-reduction parameters \cite{alb+krug}, 
$e.g.$ of the type  
of the sticking parameter in MBDLA \cite{mbdla}. More detailed predictions 
from model (\ref{ecdif})-(\ref{ecinfty}) would benefit from a 
computationally more efficient formulation. Work along this lines
is currently underway \cite{us}.

\begin{acknowledgments}
We are pleased to acknowledge discussions with and comments by J.\ Buceta, 
F.\ Dom\'{\i}nguez-Adame, A.\ Hern\'andez-Machado, E.\ Moro, 
M.A.\ Rodr\'{\i}guez, M.A.\ Rubio, A.\ S\'anchez, and L.\ V\'azquez. 
This work has been supported by DGES (Spain) grant No.\ BFM2000-0006.
\end{acknowledgments}

\bigskip

\hrule


\begin{thebibliography}{99}
\bibitem{stick} The coefficient $k_D$ has dimensions of velocity and
is related \cite{brekel+pg+brz} by kinetic theory with the sticking 
probability $\tilde{s}$ of particles diffusing near a partially absorbing 
boundary as $k_D \propto D L_{\rm mfp}^{-1} \, \tilde{s}/(2-\tilde{s})$, 
with $L_{\rm mfp}$ the mean free path for the diffusing particle 
and $D$ the diffusion constant.
\end{thebibliography}
\end{document}